\documentclass[12pt]{article}

\usepackage{latexsym}
\usepackage{amssymb,amsfonts,amsmath}
\usepackage{graphicx} 
\usepackage{indentfirst}
\usepackage{bbm}
\usepackage{amssymb}
\usepackage{verbatim}
\usepackage{amsmath, amsthm,amssymb}
\usepackage{mathrsfs}
\usepackage{hyperref}
\usepackage{amsfonts}
\usepackage{dsfont}
\usepackage{cite}

\parskip=\medskipamount

\newcommand {\cA}{{\cal A}}

\newcommand {\cI}{{\cal I}}

\newcommand {\cN}{{\cal N}}

\newcommand {\cP}{{\cal P}}

\newcommand {\cW}{{\cal W}}

\def\a{\alpha}

\def\b{\beta}

\def\d{\delta}
\def\e{\epsilon}

\def\g{\gamma}
\def\G{\Gamma}

\def\k{\kappa}
\def\l{\lambda}

\def\o{\omega}

\def\q{\theta}

\def\s{\sigma}

\def\x{\xi}
\def\z{\zeta}
\def\D{\Delta}
\def\F{\Phi}
\def\J{\Psi}
\def\L{\Lambda}

\def\P{\Pi}

\def\U{\Upsilon}

\def\rd{{\rm d}}
\def\ri{{\rm i}}

\newcommand{\ve}{\varepsilon}                            

\newcommand{\pa}{\partial}                           
\newcommand{\hf}{\frac12}

\newcommand{\be}{\begin{equation}}
\newcommand{\ee}{\end{equation}}
\newcommand{\bea}{\begin{eqnarray}}
\newcommand{\eea}{\end{eqnarray}}
\newcommand{\non}{\nonumber}
\newcommand{\1}{{\underline{1}}}
\newcommand{\2}{{\underline{2}}}

\def\double #1{#1{\hbox{\kern-2pt $#1$}}}

\newif\ifdtup

\newcommand{\bsubeq}{\begin{subequations}}
\newcommand{\esubeq}{\end{subequations}}

\numberwithin{equation}{section}

\begin{document}

\begin{titlepage}
\begin{flushright}
May, 2019 \\
\end{flushright}
\vspace{5mm}

\begin{center}
{\Large \bf 
Linearised actions for $\cal N$-extended (higher-spin) superconformal gravity}
\\ 
\end{center}

\begin{center}

{\bf
Evgeny I. Buchbinder, Daniel Hutchings, Jessica Hutomo \\
and Sergei M. Kuzenko} \\
\vspace{5mm}

\footnotesize{
{\it Department of Physics M013, The University of Western Australia\\
35 Stirling Highway, Crawley W.A. 6009, Australia}}  
~\\

\vspace{2mm}
~\\
\texttt{Email: evgeny.buchbinder@uwa.edu.au, daniel.hutchings@research.uwa.edu.au, \\
jessica.hutomo@research.uwa.edu.au, sergei.kuzenko@uwa.edu.au}
\vspace{2mm}

\end{center}

\begin{abstract}
\baselineskip=14pt
The off-shell actions for $\cal N$-extended conformal supergravity theories in three dimensions 
were formulated in \cite{BKNT-M2,KNT-M}
for $1\leq {\cal N} \leq 6$ using a universal approach. 
Each action is generated by a closed super three-form which is constructed in terms of the constrained geometry of $\cal N$-extended conformal
superspace. In this paper we initiate a program to recast these actions 
(and to formulate their higher-spin counterparts) 
in terms of unconstrained gauge prepotentials as integrals over the full superspace.
We derive transverse projection operators in
$\cal N$-extended Minkowski superspace
and then use them to construct linearised rank-$n$ super-Cotton tensors
and off-shell  $\cal N$-extended superconformal actions. We also propose 
off-shell gauge-invariant actions to describe massive higher-spin supermultiplets 
in $\cal N$-extended supersymmetry. Our analysis leads to general expressions for identically conserved higher-spin current multiplets in $\cal N$-extended supersymmetry. 
\end{abstract}
\vspace{5mm}

\vfill

\vfill
\end{titlepage}

\newpage
\renewcommand{\thefootnote}{\arabic{footnote}}
\setcounter{footnote}{0}

\tableofcontents{}
\vspace{1cm}
\bigskip\hrule

\allowdisplaybreaks


\section{Introduction}

Superprojectors \cite{SalamS,Sokatchev,IS,SG} are superspace
projection operators which single out irreducible representations of supersymmetry.  
There are various applications of such operators in the literature, 
including the constructions 
of superfield equations of motion \cite{OS1,OS2} and gauge-invariant actions
\cite{GS,GKP}. In the case of $\cN=1$ anti-de Sitter supersymmetry
in four spacetime dimensions, the superprojectors introduced in \cite{IS}
define the two types of complex linear  
supermultiplets (transverse and longitudinal)
that are used as the compensators 
in the massless supersymmetric higher-spin gauge theories 
proposed in \cite{KS94} and generalised in \cite{BHK}.\footnote{The Ivanov-Sorin superprojectors \cite{IS} have a natural 
three-dimensional analogue in the framework of $(1,1)$ anti-de Sitter supersymmetry
\cite{HKO}.}

In this paper we first derive transverse spin projection operators in three-dimensional 
$\cN$-extended Minkowski superspace,  ${\mathbb M}^{3|2\cN}$, 
and then make use of these superprojectors to construct
linearised off-shell actions for $\cN$-extended (higher-spin) conformal supergravity
in terms of unconstrained prepotentials. For the $1\leq \cN \leq 6$ cases, 
the complete nonlinear actions for $\cN$-extended conformal supergravity were
derived in \cite{BKNT-M2,KNT-M} using the off-shell formulation for 
$\cN$-extended conformal supergravity developed in \cite{BKNT-M1}.\footnote{The
$\cN=1$ and $\cN=2$ conformal supergravity theories were constructed for the first time by  van Nieuwenhuizen \cite{vN} and  
Ro\v{c}ek and  van Nieuwenhuizen  \cite{RvN}, respectively. 
The off-shell action for
$\cN=6$ conformal supergravity was independently derived by 
Nishimura and Tanii \cite{NT}.
On-shell  formulations for $\cN$-extended conformal supergravity with  $\cN>2$
were given in \cite{LR89,NG}.}
Since the complete nonlinear actions 
are known, it is natural to ask the following question: What is the point 
of constructing linearised conformal supergravity actions? The answer is that 
the supergravity actions proposed in \cite{BKNT-M2} are realised using 
 certain closed super three-forms which are constructed in terms of 
 the constrained geometry of $\cN$-extended conformal superspace \cite{BKNT-M1}.
 However, it may be shown that the constraints can be 
 solved in terms of unconstrained 
 prepotentials. Modulo purely gauge degrees of freedom,
 the structure of unconstrained conformal gauge prepotentials are as follows:
 $H_{\a\b\g} $ 
for $\cN=1$ \cite{GGRS}, $H_{\a\b}$ for $\cN=2$ \cite{ZP,Kuzenko12}, 
$H_\a$ for $\cN=3$ \cite{BKNT-M1}, and $H$ for $\cN=4$ \cite{BKNT-M1}. 
The action for conformal supergravity, $S_{\rm CSG} $, may be reformulated 
in terms of the gauge prepotential $H$ (with indices suppressed),
$S_{\rm CSG} =S_{\rm CSG} [H]$, 
and such a formulation is expected to be essential for doing quantum supergraph calculations (say, in off-shell $\cN$-extended versions of topologically massive supergravity \cite{DK,Deser})
and other applications. In order to determine the structure of $S_{\rm CSG} [H] $,
the starting point is to first construct a linearised conformal supergravity action, which is 
one of the aims of this paper.

In this paper we propose a universal approach to
construct linearised  actions for $\cN$-extended superconformal gravity theories and 
their higher-spin extensions. Our conceptual setup will be described in the next section.
Then it will be applied to theories with $1\leq \cN \leq 6$ in sections 3 to 8. 
Our main results and their implications and generalisations will be discussed in section 9.
The main body of the paper is accompanied by two technical appendices.


\section{Conceptual setup and the main results} \label{section2}

The geometry of $\cN$-extended conformal superspace \cite{BKNT-M1} 
is formulated in terms of a single curvature superfield, which is the super-Cotton 
tensor $\cW$ (with suppressed indices). This tensor is encoded in the action for 
conformal supergravity by the rule \cite{BKNT-M1,KNT-M}
\bea
\cW \propto \frac{\d S_{\rm CSG} [H]}{\d H}~.
\eea
The functional structure of $\cW$ depends on the choice of $\cN$. 
 The $\cN=1$ super-Cotton tensor \cite{KT-M12}
 is a primary symmetric rank-3 spinor superfield
 $\cW_{\a\b\g}$ of dimension 5/2, which obeys the conformally invariant constraint
 \cite{BKNT-M1}
\bea
\nabla^\a \cW_{\a \b\g} = 0 ~. 
\label{WW2.2}
\eea
In the $\cN=2$ case,  the super-Cotton tensor \cite{ZP,Kuzenko12} 
is a primary symmetric rank-2 spinor superfield 
 $\cW_{\a\b}$ of dimension 2, which obeys the Bianchi identity  \cite{BKNT-M1}
\bea
\nabla^{\a I} \cW_{\a\b} = 0 ~. 
\label{WW2.3}
\eea
In the $\cN=3$ case,  the super-Cotton tensor is a  primary spinor superfield
 $\cW_{\a}$ of dimension 3/2 constrained by \cite{BKNT-M1}
 \bea
\nabla^{\a I} \cW_\a = 0 ~.
\label{WW2.4}
\eea
In the $\cN=4$ case, the super-Cotton tensor is a primary  scalar
 superfield  $\cW$ of dimension 1
constrained by \cite{BKNT-M1}
\bea
\nabla^{\a I}\nabla_{\a}^J \cW=\frac{1}{4}\d^{IJ}\nabla^{\a K}\nabla_{\a}^K \cW~.
\label{WW2.5}
\eea
For $\cN>4$,  the super-Cotton tensor \cite{HIPT,KLT-M11}
is a completely antisymmetric tensor $\cW^{IJKL}$
of dimension 1 constrained by \cite{BKNT-M1}
\bea \nabla_{\a}^I \cW^{JKLP} = \nabla_\a^{[I} \cW^{JKLP]} 
- \frac{4}{\cN - 3} \nabla^Q_{\a} \cW^{Q [JKL} \d^{P] I} \ .
\label{2.666}
\eea
In the above relations, $ \nabla^I_\a$ denotes the spinor covariant derivative
of $\cN$-extended conformal superspace \cite{BKNT-M1}.

The above consideration implies that we need expressions for linearised super-Cotton 
tensors  in terms of the gauge prepotentials, $W=W(H)$, in order to obtain 
linearised conformal supergravity actions for $1\leq \cN \leq 4$. We will consider a more 
general problem and work out linearised rank-$n$ super-Cotton tensors 
$W_{\a(n)} (H)$ as descendants of superconformal gauge prepotentials
$H_{\a(n)}$ in the case of $\cN$-extended supersymmetry.

In this paper we make use of the notation and conventions adopted in \cite{KPT-MvU}.
In particular,  $\cN$-extended Minkowski superspace ${\mathbb M}^{3|2\cN}$ 
is parametrised by  
real coordinates $z^A= (x^a, \theta^{\alpha}_I)$, where 
the $R$-symmetry  index of $\theta^{\alpha}_I$
takes $\cN$ values, $I=  {\1} , {\2} , \dots , {\underline{\cN}} $.\footnote{Since the 
$R$-symmetry group is ${\rm SO}(\cN)$, and the corresponding 
 indices are raised and lowered using the Kronecker delta, we do not distinguish 
 between upper and lower ${\rm SO}(\cN)$ indices.}
The spinor covariant derivatives $D^I_\a$ obey the anti-commutation relation
\bea
\{ D^{I}_{\alpha}, D^{J}_{\beta}\} =2 \ri\, \d^{IJ} \partial_{\alpha \beta}~.
\label{1.1}
\eea
An important role in our analysis will be played by 
the operator 
\bea
\D = -\frac{\ri}{2\cN} D^{\a I } D^I_\a 
\label{1.2}
\eea
with the following properties
\begin{subequations}\label{1.3} 
\bea
D^{\a I} \J_\a &=& 0 \quad \implies \quad \D \J_\a = \pa_\a{}^\b \J_\b~,
\label{1.3a} 
\\
D^{\a I} \J_\a &=& 0 \quad \implies \quad D^{\a I} \D \J_\a = 0~.
\label{1.3b} 
\eea
\end{subequations}
Another important property of $\D$ is
\bea
\big[\Delta \,, D^{\b I}D^{I}_{\a} \big] &=& 0~.
\label{1.4}
\eea

As a generalisation of the earlier $\cN=1$ \cite{K16,KT} and $\cN=2$ \cite{KO} results, 
in this paper we propose a superconformal gauge-invariant action of the form\footnote{The 
functional structure of \eqref{action} is reminiscent of the conformal higher-spin actions 
in four dimensions \cite{FT,FL-4D}.}
\bea
{S}^{(n|\cN)} [  H_{\a(n)}] = \frac{ \ri^n}{2} 
   \int \rd^{3|2\cN}z \, H^{\a(n)} 
{W}_{\a(n)}\big(H\big) ~, \qquad n>0~,
\label{action}
\eea
where the dynamical superfield
$H_{\a(n)}= H_{\a_1 \dots \a_n}= H_{(\a_1 \dots \a_n)}$
is a real  symmetric rank-$n$ spinor which is defined modulo 
gauge transformations
\bea
\d_\z H_{\a(n )} &=& \ri^n D^I_{(\a_1 } \z^I_{\a_2 \dots \a_n)} ~.
\label{1.6}
\eea
The field strength $W_{\a(n)}$ in \eqref{action} is a local descendant of $H_{\a(n)}$.
It is a real completely symmetric 
rank-$n$ spinor, which is required to obey several conditions:
\begin{enumerate}
\item $W_{\a(n)}$ is gauge invariant, 
\bea
{W}_{\a(n) }\big(\d_\z H \big) =0~; \label{1.7}
\eea
\item $W_{\a(n)}$ is transverse,
\bea
D^{\b I} W_{\b \a_1 \dots \a_{n-1}} =0~; \label{1.8}
\eea
\item $W_{\a(n)}$ is a primary superconformal multiplet
(all relevant technical details about the $\cN$-extended superconformal 
group are collected in Appendix \ref{appendixA}).
The condition \eqref{1.8} uniquely fixes the dimension $d_{W_{\a(n)}}$
of ${W}_{\a(n) }$.
\end{enumerate}
The dimension $d_{H_{\a(n)}}$
of ${H}_{\a(n) }$ is also fixed uniquely if we require $H_{\a(n)}$ and the 
gauge parameter $\z_{\a(n-1)}$ in \eqref{1.6} to be superconformal primary.
The dimensions are: 
\bea
d_{H_{\a(n)}} = 2-\cN -\frac n2~, \qquad
d_{W_{\a(n)}} = 1 +\frac n2~.
\label{1.9}
\eea

In this paper we will demonstrate that the above conditions determine
$W_{\a(n)}$, modulo an overall numerical factor, in the form
\bea
W_{\a(n)} \big(H\big) =  \D^{n+\cN-1}  \P^{\perp}_{[n]}H_{\a(n)}~,
\label{2.16}
\eea
where $ \P^{\perp}_{[n]}$ is  a transverse projector, 
\bea
 \P_{[n]}^{\perp}  \P_{[n]}^{\perp} =  \P_{[n]}^{\perp}~.
 \label{idempotent}
 \eea
 By definition, the projection operator $\P^{\perp}_{[n]}$ acts on the space of real symmetric  
rank-$n$ spinors $\J_{\a(n)}= \J_{\a_1 \dots \a_n}= \J_{(\a_1 \dots \a_n)}$ 
by the rule
\bea
\P^{\perp}_{[n]} \J_{\a(n)} := \P_{\a_1}{}^{\b_1} \dots \P_{\a_n} {}^{\b_n} 
\J_{\b_1 \dots \b_n}\equiv \J^{\perp}_{\a_1 \dots \a_n}
=\J^{\perp}_{(\a_1 \dots \a_n)}~,
\label{2.18}
\eea 
where the operator $\P_\a{}^\b$ has the following universal properties
\bsubeq\label{5.2}
\bea
D^{\a I} \P_{\a}{}^{\b} &=& 0~,  \label{5.2a}\\
\P_{\a}{}^{\b} D^{I}_{\b}&=& 0~, \label{5.2b}\\
\P_{\a}{}^{\b}\P_{\b}{}^{\g} &=& \P_{\a}{}^{\g}~, \label{5.2c}\\
\big[ \P_{\a}{}^{\b}  , \P_{\g}{}^{\d} \big]&=&0~. \label{5.2d}
\eea
\esubeq
The superfield $\J^{\perp}_{\a_1 \dots \a_n}$ defined by \eqref{2.18}
 is completely symmetric, as a consequence of 
the identity \eqref{5.2d}, and is transverse, 
\bea
D^{I \b}  \P^{\perp}_{[n]}  \J_{\b \a_1 \dots \a_{n-1}} =0 ~.
\label{Tpro}
\eea
In general,  a real symmetric rank-$n$ spinor superfield $T_{\a(n)} $ is called
 transverse (or divergenceless) 
if it obeys the constraint \eqref{1.8}.
It holds that\footnote{For $n>1$ the spinor  transverse condition \eqref{2.20} implies 
that $T_{\a(n)}$ is transverse in the usual sense, that is $\pa^{\b\g} T_{\b\g \a(n-2)} =0$.
}
\bea
D^{I \b} T_{\b \a_1 \dots \a_{n-1}} =0 \quad \implies \quad 
 \P^{\perp}_{[n]} T_{\a(n)}  = T_{\a(n)} ~.
 \label{2.20}
\eea
In the case of $\cN=2$ supersymmetry, one can define the
 so-called {\it complex transverse linear} superfields \cite{KO}. 
They obey a weaker constraint than \eqref{2.20}, see section \ref{section4}.

One of the main goals of this paper is the explicit construction of $ \P_\a{}^\b$
for different supersymmetry types,   $1\leq \cN \leq 6$. 
Our ansatz  for $ \P_\a{}^\b$ is
\bea
 \P_\a{}^\b =D^{\b I}D^{I}_{\a} \, F \big( \D , \Box\big)~,
 \eea
 for some function $ F \big( \D , \Box\big)$ to be determined.
 Due to \eqref{1.4} and the identity
 \bea
D^{\b I}D^{I}_{\a} &=&\ri  \cN  \big(\pa_{\a}{}^{\b} + \d_{\a}{}^{\b}\Delta \big)~,
\label{2.22}
\eea
the condition \eqref{5.2d} is satisfied.

This work is a natural 
continuation of the research described in the non-supersymmetric case in \cite{BKLFP}.


\section{${\cal N}=1$ supersymmetry}\label{section3}

Let $D_\a$ be the spinor covariant derivative of  $\cN=1$ Minkowski superspace. 
Making use of \eqref{1.1} allows us to obtain a number of useful identities including the following:
\bsubeq\label{3.3}
\bea
 D_{\a } D_{\b} &=& \ri \pa_{\a \b} +\frac{1}{2} \ve_{\a \b} D^2~,\\
D^\a D_\b D_\a &=& 0 \quad \implies \quad [D_{\a} D_{\b}, D_{\g} D_{\d}]=0~, \label{3.3b} \\
D^2 D_{\a} &=& -D_{\a} D^2 = 2\ri \pa_{\a\b}D^{\b}~, \label{3.3c}\\
D^2 D^2 &=& -4 \Box ~, \label{3.3d}
\eea
\esubeq
where we have denoted  $D^2= D^{\a} D_{\a}$ and 
$\Box = \pa^{a} \pa_{a} = -\hf \pa^{\a \b} \pa_{\a \b}$~.


\subsection{Superprojectors}

Let us consider the following operator 
\bea
\P_{\a}{}^{\b} = -\frac{D^2}{4 \Box} 
D^{\b}D_{\a} 
= -\frac{\ri}{2} \frac{\D}{ \Box} 
D^{\b}D_{\a} 
~,
\label{2.2}
\eea
which acts on the space of real spinor superfields, 
$\J_\a \to \P_\a{}^\b \J_\b$.
It satisfies the projector property \eqref{5.2c},
as a consequence of \eqref{3.3}.
The identities \eqref{3.3b} and \eqref{3.3c} imply that
it also satisfies the conditions \eqref{5.2a} and \eqref{5.2b}.
If $\J_\a$ is transverse, $D^\a \J_\a=0$, it holds that 
\bea
D^\a \J_\a=0 \quad \implies \quad \P_{\a}{}^{\b} \J_{\b} =\J_\a~.
\eea
We conclude that $\P_{\a}\,^{\b}$ is the projection operator onto the space of transverse 
spinor superfields and thus $\P_{\a}\,^{\b}$ can be called a transverse projector.

As a higher-rank generalisation of  \eqref{2.2} we introduce 
a  projection operator $\P^{\perp}_{[n]}$ which
acts on the space of real symmetric  rank-$n$ spinors
$\J_{\a(n)}= \J_{\a_1 \dots \a_n}= \J_{(\a_1 \dots \a_n)}$.
It is defined by the rule \eqref{2.18}.
The superfield $\J^{\perp}_{\a_1 \dots \a_n}$ defined by \eqref{2.18} 
 is completely symmetric as a consequence of 
the identity  \eqref{3.3b}. The same identity implies 
that 
\bea
D^{\b} \P^{\perp}_{[n]} \J_{\b \a_1 \dots \a_{n-1}}=0~,
\eea
and therefore $\P^{\perp}_{[n]} \J_{\a(n)} $ is transverse for every
superfield $ \J_{\a(n)}$.
It is not difficult to see that  $\P^{\perp}_{[n]}  $ maps every transverse superfield to itself, 
 \bea
 D^{\b}  \J_{\b \a_1 \dots \a_{n-1}}=0 \quad \implies \quad 
 \P^{\perp}_{[n]} \J_{\a(n)} = \J_{\a(n)} ~.
 \eea
We conclude that $\P^{\perp}_{[n]} $ is the projector onto the space of transverse
rank-$n$ spinor superfields.

We now turn to studying the $\cN=1$ projection 
operator $\P^{\parallel}_{[n]} := \mathbbm{1}_{[n]} -  \P_{[n]}^{\perp}$.
Given an arbitrary symmetric real rank-$n$ spinor $\J_{\a(n)}$, we 
obtain
\bsubeq
\bea
\big( \mathbbm{1}_{[n]} -  \P_{[n]}^{\perp}\big) \J_{\a(n)}
&=& \ri^n D_{(\a_1} \l_{\a_{2} \dots \a_{n})}~,
\eea
where we have denoted
\bea
\l_{\a_1 \dots \a_{n-1}}&:=& -(-\ri)^n\sum_{j=1}^{n} \frac{1}{(4\Box)^j} \binom{n}{j} 
(D^2)^j
D^{\b_{n-1}}  D_{(\a_{n-1}} \dots D^{\b_{n-j+1}}  D_{\a_{n-j+1}}
\non\\
&&\times D^{\b_n}\J_{\a_{1} \dots \a_{n-j}) \b_{n-j+1} \dots \b_n}~.
\eea
\label{3.12}
\esubeq
In order to prove \eqref{3.12}, it is useful to rewrite $\P_{\a}\,^{\b}$ in  the form
\bea
\P_{\a}\,^{\b} = \d_{\a}\,^{\b}- \frac{D^2}{4\Box}D_{\a}D^{\b}~.
\eea

Any symmetric rank-$n$ spinor  of the form 
$\F_{\a(n)} =D_{(\a_1} \U_{\a_{2} \dots \a_{n})}$ is said to be longitudinal.
The  projector $\P^{\parallel}_{[n]} $ maps every longitudinal superfield 
to itself, 
$\big( \mathbbm{1}_{[n]} -  \P_{[n]}^{\perp}\big) \F_{\a(n)} = \F_{\a(n)}$.
Thus $\P^{\parallel}_{[n]} $ is the projector onto the space of
longitudinal
rank-$n$ spinor superfields.


\subsection{Linearised rank-$n$ super-Cotton tensor}

Given a positive integer $n$, the rank-$n$ 
super-Cotton tensor \cite{K16} (see also \cite{KT,KP}) is 
\bea
W_{\a(n)} (H)= \Big(-\frac{\ri}{2}\Big)^n D^{\b_1}D_{\a_1} \dots D^{\b_n} D_{\a_n}
H_{\b_1 \dots \b_n}~.
\label{N1CT}
\eea
Its fundamental properties are the following: 
(i) it is invariant under the gauge transformations \eqref{1.6};
and (ii) it obeys the conservation identity \eqref{1.8}.

The choice  $n=1$  in  \eqref{N1CT} 
corresponds to the gauge-invariant field strength
of an Abelian vector multiplet   \cite{Siegel}. 
The case $n=2$ corresponds to the super-Cottino tensor \cite{K16} 
which is the gauge-invariant field strength of a superconformal gravitino 
multiplet.\footnote{Among the component fields of $W_{\a\b}$ is the so-called 
Cottino tensor $C_{\a\b\g} =C_{(\a\b\g)}$, which is
 the gauge-invariant field strength of a conformal gravitino 
\cite{DK,GPS,ABdeRST}.
}
Choosing $n=3$ in \eqref{N1CT} gives the linearised version 
of the $\cN=1$ super-Cotton tensor \cite{KT-M12}. 
Finally, for $n>3$ the component fields of $W_{\a(n)}$ contain 
 linearised bosonic \cite{PopeTownsend} and fermionic \cite{K16}
 higher-spin Cotton tensors.

The super-Cotton tensor \eqref{N1CT} can be expressed in terms of the 
transverse projection operator $ \P_{[n]}^{\perp}$ in the form
\bea
W_{\a(n)} =  \D^n  \P^{\perp}_{[n]}H_{\a(n)}~,
\label{3.99}
\eea
which is a special case of \eqref{2.16}. 
In order to demonstrate  that \eqref{3.99} is equivalent to \eqref{N1CT}, it suffices to note that, 
in accordance with \eqref{3.3d}, $\D$ is a square root of the d'Alembertian,
\bea
\D^2 = \Box~.
\label{3.100}
\eea
This property allows us to obtain alternative expressions 
for $W_{\a(n)}$, depending on whether an explicit value of $n$ is even or odd.
These expressions are:
\begin{subequations}
\bea
W_{\a(2s)} &=& \Box^{s}  \P^{\perp}_{[2s]}H_{\a(2s)}~, \qquad s= 1, 2, \dots\\
W_{\a(2s+1)} &=& 
\Box^{s}\D  \P^{\perp}_{[2s+1]}H_{\a(2s+1)} ~, \qquad s=0, 1,\dots 
\eea
\end{subequations}


\section{${\cal N}=2$ supersymmetry}\label{section4}

In the case of $\cN=2$ supersymmetry, it is often useful to work with 
a complex basis for the spinor covariant derivatives. 
Such a basis is introduced \cite{KPT-MvU} by replacing 
the real covariant derivatives  $D^{I}_{\a} = (D^{\1}_{\a}, D^{\2}_{\a})$ 
with the complex operators $D_{\a}$ and $\bar D_{\a}$ defined by:
\bea
D_{\a} = \frac{1}{\sqrt{2}}(D^{\1}_{\a} - \ri D^{\2}_{\a})~, \qquad \bar D_{\a} = -\frac{1}{\sqrt{2}}(D^{\1}_{\a} + \ri D^{\2}_{\a})~.
\label{D2real}
\eea
As follows from \eqref{1.1}, 
the complex spinor covariant derivatives 
satisfy the anti-commutation relations 
\bea
\{D_{\alpha}, D_{\beta}\}=0\,, \qquad
\{\bar D_{\alpha}, \bar D_{\beta}\}=0\,, \qquad
\{ D_{\alpha}, \bar D_{\beta}\} =-2 \ri \partial_{\alpha \beta}\,.
\label{3.2}
\eea
In terms of the new covariant derivatives, 
one naturally defines important off-shell supermultiplets including 
(i)  a chiral superfield $\F$ constrained by $\bar D_\a \F=0$; 
(ii) a complex linear superfield $\G$ constrained by $\bar D^2 \G =0$; and
(iii) a real linear superfield $L=\bar L$ constrained by $\bar D^2 L =0$.


\subsection{Superprojectors in the complex basis}

We introduce the operator
\bsubeq
\bea
\P_{\a}{}^{\b} = \frac{\ri}{4 \Box} {\Delta}\big( \bar D^{\b} D_{\a} + D^{\b} \bar D_{\a}\big)~,
\eea
where $\D$ denotes the $\cN=2$  version of  \eqref{1.2} written in the complex basis,
\bea
{\Delta} &=& \frac{\ri}{2} D^{\a} \bar{D}_{\a} = \frac{\ri}{2} \bar D^{\a} D_{\a}~.
\eea
\label{3.3a}
\esubeq
Making use of the anti-commutation relations \eqref{3.2}, 
 $\P_{\a}{}^{\b}$ can be rewritten in the form
\bea
\P_{\a}{}^{\b} = \frac{\Delta}{2\Box} \big( \pa_{\a}\,^{\b} + \d_{\a}\,^{\b} {\Delta} \big)~,
\label{3.4}
\eea
which implies the validity of \eqref{5.2d}.
One may also check that $\P_{\a}{}^{\b} $ satisfies the other three conditions 
in \eqref{5.2}, keeping in mind that $D^I_\a$ stands for $D_\a$ and $\bar D_\a$.
For this it suffices to make use of identities of the type
$\bar D^{\b} {\Delta} = \frac{\ri}{4} \bar D^2 D^{\b}$.
Thus $\P_{\a}{}^{\b}$ is the  projection operator 
onto the space of transverse spinor superfields.

Given the projector $\P_\a{}^\b$, we define
the transverse projection operator $\P^{\perp}_{[n]} $ by the rule \eqref{2.18}.
It acts on the space of real symmetric rank-$n$ spinors, $\J_{\a(n)}$,
and projects it onto the subspace of transverse superfields such that 
\bea
\bar D^{\b} \P^{\perp}_{[n]}
\J_{\a_1 \dots \a_{n-1}\b}= D^{\b} \P^{\perp}_{[n]}
\J_{\a_1 \dots \a_{n-1}\b} =0~.
\eea

At this point it is worth pausing in order to make a few comments. 
According to the terminology of \cite{KO}, 
every real symmetric rank-$n$ spinor $T_{\a(n)}= \bar T_{\a(n)}$ constrained by
\bea
\bar D^{\b} T_{\a_1 \dots \a_{n-1}\b} =0 \quad \Longleftrightarrow \quad
D^{\b}
T_{\a_1 \dots \a_{n-1}\b} =0
\label{4.60}
\eea
is said to be {\it real transverse linear}. It is linear since the above constraint implies
\bea
 \bar D^{2}  T_{\a(n)}=0 \quad \Longleftrightarrow \quad D^{2} T_{\a(n)}=0~,
\eea
as a consequence of  \eqref{3.2}. There also exist 
{\it complex transverse linear} superfields $\G_{\a(n)}$ which obey the only constraint 
\bea
\bar D^{\b} \G_{\a_1 \dots \a_{n-1}\b} =0 \quad \implies \quad
\bar D^2 \G_{\a(n)} =0~.
\label{4.88}
\eea
The general solution to this constraint proves to be 
\bea
\G_{\a (n) } &=& \bar D^\b \x_{\b \a_1 \dots \a_n } ~, 
\eea
with the prepotential $\x_{\a(n+1)}$ being  complex unconstrained.
A general solution to \eqref{4.60} is more complicated, and  will be discussed below.

In the case of $\cN=1$ supersymmetry, $\D$ is an invertible operator, eq. \eqref{3.100}.
This is no longer true for $\cN=2$, in particular  due to the relation 
\bea
\Box = \D^2 + \frac{1}{16} \big\{ \bar D^2 , D^2 \big\}~,
\label{Salam}
\eea
which is the three-dimensional analogue of a famous result in four dimensions \cite{SalamS}.
This relation can be rewritten in the form 
\bea
{\mathbbm 1}& =& \cP_{(\ell)} + \cP_{(+)} +\cP_{(-)}~, \non \\
\cP_{(\ell)} &=& \frac{1}{\Box} \D^2~, 
\qquad \cP_{(+)}=\frac{1}{16\Box}  \bar D^2 D^2 ~, \qquad 
\cP_{(-)}=\frac{1}{16\Box}  D^2 \bar D^2 ~,
\label{4.100}
\eea
where $\cP_i =\big(\cP_{(\ell)}, \cP_{(+)},\cP_{(-)}\big)$ are orthogonal projectors,
$\cP_i \cP_j = \d_{ij} \cP_i$. The operator $\cP_{(\ell)}$ is the projection operator 
onto the space of real linear superfields,
\bea
\bar D^2 \cP_{(\ell)} V = D^2 \cP_{(\ell)} V =0~,
\eea
for every real scalar $V$. The operator $\cP_{(+)}$ is the projection operator 
onto the space of chiral superfields,
\bea
\bar D_\a \cP_{(+)} U = 0~.
\eea
Finally, the operator $\cP_{(-)}$ is the projection operator 
onto the space of antichiral superfields.

We now turn to studying the $\cN=2$ projection 
operator $\P^{\parallel}_{[n]} := \mathbbm{1}_{[n]} -  \P_{[n]}^{\perp}$.
For an arbitrary real symmetric rank-$n$ superfield $H_{\a(n)}$, 
we obtain 
\bsubeq\label{longH}
\bea
 \big( \mathbbm{1} -  \P^{\perp}\big) H_{\a(n)}
&=& \bar D_{(\a_n} \L_{\a_{1} \dots \a_{n-1})} 
- (-1)^n D_{(\a_n} \bar \L_{\a_{1} \dots \a_{n-1})}~,
\eea
where we have denoted 
\bea
\L_{\a_1 \dots \a_{n-1}} &=& 
- \sum_{j=1}^{n} \bigg(\frac{\ri}{4\Box}\bigg)^j \binom{n}{j} 
D^{\b_n} A_{(\a_{n-1}}^{\quad \b_{n-1}} \dots A_{\a_{n-j+1}}^{\quad \b_{n-j+1}}
\Delta^{j}H_{\a_{1} \dots \a_{n-j}) \b_{n-j+1} \dots \b_n} \non\\
&&+ \frac{1}{8\Box} D^{\b} \bar D^2 H_{\b \a_1 \dots \a_{n-1}}~.
\eea
\esubeq
Here the operator $A_{\a}{}^{\b}$ is defined by
\bea
A_{\a}{}^{\b} := D_{\a} \bar D^{\b} + \bar D_{\a} D^{\b}
\eea
and satisfies the  property
\bea
[\Delta, A_{\a}\,^{\b}]=0~,
\eea
which is crucial for our analysis.
In order to prove the relation \eqref{longH}, it is useful to rewrite $\P_{\a}\,^{\b}$ in  the form
\bea
\P_{\a}{}^{\b} = \d_{\a}{}^{\b} -\d_\a{}^\b \big(  \cP_{(+)} +  \cP_{(-)} \big)
+ \frac{\ri}{4\Box}\Delta A_{\a}\,^{\b}~,
\eea
where $ \cP_{(+)}$ and $ \cP_{(-)} $ are the chiral and antichiral projection operators 
\eqref{4.100}.
It is natural to call the operator 
$\P^{\parallel}_{[n]} := \mathbbm{1}_{[n]} -  \P_{[n]}^{\perp}$ a longitudinal superprojector.

The relation \eqref{longH} naturally leads to the gauge transformation law
\bea
\d H_{\a(n)} = g_{\a(n)} + \bar g_{\a(n)}~, \qquad 
g_{\a(n)} = \bar{D}_{(\a_{1}}L_{\a_{2}...\a_{n})} ~,
\label{4.1888}
\eea
with the gauge parameter $L_{\a(n-1)}$ being  complex unconstrained. 
This transformation law was postulated in \cite{KO} to describe the gauge freedom 
of a superconformal gauge prepotential $H_{\a(n)}$. The parameter $g_{\a(n)}$ in 
\eqref{4.1888} is an example of a {\it complex longitudinal linear} superfield \cite{KO}.
In general, such a superfield $G_{\a(n)}$ is constrained by 
\bea
\bar D_{(\a_1 } G_{\a_2 \dots \a_{n+1})} =0 \quad \implies \quad \bar D^2 G_{\a(n)} =0~.
\eea
This constraint can be compared with \eqref{4.88}. For $n=0$ this constraint is equivalent to the chirality condition.


\subsection{Linearised rank-$n$ super-Cotton tensor}

In this subsection we derive a new representation for the 
linearised  ${\cal N}=2$ rank-$n$ super-Cotton tensor, with $n > 1$.
This real tensor superfield is a descendant of the superconformal gauge superfield $H_{\a(n)}$, 
which was constructed in  \cite{KO} in the form
\bea
W_{\a(n)} (H) &:=& \frac{1}{2^{n-2}}  \sum_{j=0}^{\lfloor\frac{n}{2}\rfloor} \Bigg\{\binom{n}{2j} {\Delta} \Box^j \pa_{(\a_1}^{\,\,\,\b_1} \dots \pa_{\a_{n-2j}}^{\,\,\,\b_{n-2j}} H_{\a_{n-2j+1} \dots \a_{n)} \b_{1} \dots \b_{n-2j}}
\non \\
&&+ \binom{n}{2j+1} {\Delta}^2 \Box^j \pa_{(\a_1}^{\,\,\,\b_1} \dots \pa_{\a_{n-2j-1}}^{\,\,\,\b_{n-2j-1}} H_{\a_{n-2j} \dots \a_{n)} \b_{1} \dots \b_{n-2j-1}} \Bigg\}~. 
\label{CT}
\eea
The fundamental properties of $W_{\a(n)}(H) $ are: (i) it is invariant under the gauge transformations \eqref{4.1888}; and (ii) it is transverse, 
\bea
D^{\b}W_{\b \a_1 \dots \a_{n-1}}= \bar D^{\b}W_{\b \a_1 \dots \a_{n-1}}= 0~.
\eea
The case $n=1$ corresponds to the super-Cottino tensor \cite{K16} 
which is the gauge-invariant field strength of a superconformal gravitino multiplet.
The choice $n=2$ gives the linearised version 
of the $\cN=2$ super-Cotton tensor \cite{ZP,Kuzenko12}.

Making use of the representation \eqref{3.4}, one may show that $W_{\a(n)} (H)$ can be constructed using the transverse superprojectors $ \P_{[n]}^{\perp}$ 
in the form
\bea
W_{\a(n)} (H) = {\Delta}^{n+1}  \P_{[n]}^{\perp}H_{\a(n)}~,
\label{4.188}
\eea
which is a special case of \eqref{2.16}.
In deriving \eqref{4.188}, 
 we have made use of the special properties of the operator $\Delta$:
\bea
\Delta^{2k}= \Box^{k-1} \Delta^2~, \qquad
\Delta^{2k+1}= \Box^{k} \Delta~, \qquad k = 1,2, \dots
\label{delta2}
\eea


\subsection{Superprojectors and super-Cotton tensors in the real basis}

In the real basis for the spinor covariant derivatives, 
our transverse superprojector \eqref{3.3a} takes the form
\bea
\P_{\a}{}^{\b}= -\frac{\ri \Delta}{4\Box} D^{\b I}D^{I}_{\a}~.
\label{N2real}
\eea
It satisfies all the properties \eqref{5.2}.

We now derive another representation for  the rank-$n$  super-Cotton tensor \eqref{CT}
using the transverse superprojector $\P^{\perp}_{[n]}$ formulated in the real basis. 
Direct calculations give
\bea
W_{\a(n)} \big(H\big) =  \D^{n+1} \, \P^{\perp}_{[n]}H_{\a(n)}~,
\label{W2}
\eea
which is a special case of \eqref{2.16}.
Making use of the definition \eqref{N2real} and the property \eqref{1.4}, 
the expression \eqref{W2} turns into
\bea
W_{\a(n)}(H) = \Big(-\frac{\ri}{4 \Box}\Big)^n \Delta^{2n+1} D^{\b_1 I_1}D^{I_1}_{\a_1} \dots 
D^{\b_n I_n} D^{I_n}_{\a_n} H_{\b_1 \dots \b_n}~.
\label{W2.1}
\eea
It can be shown that the above is equivalent to
\bea
W_{\a(n)}(H) = \Big(-\frac{\ri}{4} \Big)^n \D  D^{\b_1 I_1}D^{I_1}_{\a_1} \dots D^{\b_n I_n} 
D^{I_n}_{\a_n} H_{\b_1 \dots \b_n}~,
\label{W2.2}
\eea
where we have made use of the property \eqref{delta2}.
The representation \eqref{W2.2} for the rank-$n$ super-Cotton tensor 
is clearly much  simpler than the  expression \eqref{CT} originally given in \cite{KO}.


\section{${\cal N}=3$ supersymmetry} \label{section5}

In order to construct superprojectors in the case of ${\cal N} \geq 3$ supersymmetry, we will only make use of the real basis for the spinor covariant derivatives, which satisfy the anti-commutation relation \eqref{1.1}.


\subsection{Superprojectors}

Our ${\cal N}=3$ superprojector  $\P_{\a}{}^{\b}$ is given by the  operator
\bea
\P_{\a}{}^{\b} &=& -\frac{\ri \Delta}{48\Box^2} D^{\b I} D_{\a}^{I}\big(9 \Delta^2 - \Box \big)~, 
\label{4.1}
\eea
which acts on the space of real spinor superfields.
It is possible to show that the projector satisfies the properties
\eqref{5.2}.
These properties can be proved using \eqref{1.4} and \eqref{2.22}, 
in conjunction with the following identity:
\bea
\Delta^4 &=& \frac{1}{9} \Big\{10 \Box \Delta^2 - \Box^2 \Big\}~. \label{4.3d}
\eea
It should be pointed out that in order to prove \eqref{5.2a}, we recall that the operator $\Delta$ preserves the transversality condition \eqref{1.3b}. Thus, it suffices to show that the following relation holds
\bea
D^{\a J}\Big\{D^{\b I}D^{I}_{\a}\big( 9 \Delta^2 - \Box \big)\Big\} =0~.
\label{4.4}
\eea


\subsection{Linearised rank-$n$ super-Cotton tensor}

A linearised version $W_\a(H)$ of the $\cN=3$ super-Cotton tensor \cite{BKNT-M1}
has never been computed.
Our goal in this subsection is to construct $W_\a(H)$
 and its higher spin extension $W_{\a(n)} (H)$
 using the transverse superprojector $\P^{\perp}_{[n]}$.

In accordance with \eqref{1.9}, the dimensions of the ${\cal N}=3$ gauge prepotential $H_{\a(n)}$ and its corresponding field strength $W_{\a(n)}(H)$ are $(-1-\frac{n}{2})$ and $(1+\frac{n}{2})$, respectively. 
The dimensional analysis \eqref{1.9} and the conditions \eqref{1.7} and \eqref{1.8} imply that the field strength $W_{\a(n)}(H)$ is fixed, modulo an overall constant, in the form
\bea
W_{\a(n)} \big(H\big) =  \D^{n+2} \, \P^{\perp}_{[n]}H_{\a(n)}~,
\label{W3}
\eea
which is a special case of \eqref{2.16}.
When $W_{\a(n)}$ is represented in the form \eqref{W3}, both conditions \eqref{1.7} and \eqref{1.8} are made manifest as a consequence of  \eqref{1.3b}, \eqref{5.2} and  \eqref{Tpro}.

Making use of \eqref{4.1} and the property \eqref{1.4}, the expression \eqref{W3} turns into
\bea
W_{\a(n)}(H) = \Big(-\frac{\ri}{48 \Box^2}\Big)^n \Delta^{2(n+1)} \big(9 \Delta^2 -\Box \big)^n D^{\b_1 I_1}D^{I_1}_{\a_1} \dots D^{\b_n I_n} D^{I_n}_{\a_n} H_{\b_1 \dots \b_n}~.
\label{4.7}
\eea
The expression \eqref{4.7} contains $\Box^{2n}$ in the denominator. 
However, it is possible to simplify it by making use of the following identities 
which can be derived from \eqref{4.3d}
\bsubeq
\bea
(9 \Delta^2 - \Box)^n &=& (8\Box)^{n-1} (9 \Delta^2 - \Box)~,  \label{4.6a}\\
\Delta^{2n}( 9 \Delta^2 - \Box) &=& \Box^n (9 \Delta^2 - \Box)~, \qquad n=1,2,\dots
 \label{4.6b}
\eea
\esubeq
Then, it follows from \eqref{4.6a} and \eqref{4.6b} that \eqref{4.7} is equivalent to
\bea
W_{\a(n)}(H) = \frac{1}{8}\Big(-\frac{\ri}{6} \Big)^n \big(9 \Delta^2 - \Box\big) D^{\b_1 I_1}D^{I_1}_{\a_1} \dots D^{\b_n I_n} D^{I_n}_{\a_n} H_{\b_1 \dots \b_n}~.
\label{4.10}
\eea
In the $n=1$ case, the field strength $W_{\a}$ corresponds to the linearised version of
 the ${\cal N}=3$ super-Cotton tensor. Thus, the field strength \eqref{W3} (or equivalently \eqref{4.10}) can be referred to as the rank-$n$ super-Cotton tensor. 


\subsection{Superconformal gravitino multiplet}

Let us point out that \eqref{4.6a} implies that the following operator 
\bea
\cP = \frac{1}{8\Box}(9\D^2 -\Box)
\eea
is a projector, $ \cP^2 = \cP$. This projector is relevant in the context of  a superconformal gravitino multiplet. 

In accordance with the analysis of $\cN=3$ supermultiplets of conserved currents  \cite{BKS2}, the superconformal gravitino multiplet should be described by a real scalar 
gauge prepotential $H$ of dimension $-1$, which is defined modulo 
under gauge transformations 
\bea
\d_\z H = \ri D^{\a I} D^{J}_{\a} \z^{IJ} ~, 
\qquad \z^{IJ} = \z^{JI}~, \quad \z^{II}=0~.
\eea
Associated with $H$ is a primary descendant $W(H) $ of dimension $+1$, which 
has the following properties: (i) it is gauge invariant, 
\bea
W(\d_\z H)=0~;
\eea
and (ii) it obeys the constraint 
\bea
(D^{\a I} D^{J}_{\a} - 2 \ri \d^{IJ} \Delta)W =0~.
\label{511}
\eea
We normalise this super-Cottino as
\bea
W= \frac{1}{8}(9\D^2 -\Box)H~.
\label{512}
\eea
Let us also note that acting with $D^{\b J}$ on the constraint \eqref{511} leads to
\bea
D^{\a I}\Big\{D^{\b J}D^{J}_{\a}\big( 9 \Delta^2 - \Box \big)\Big\} =0~,
\label{513}
\eea
which is the transverse condition \eqref{4.4}. In deriving \eqref{513}, we have made use of \eqref{1.1}, \eqref{2.22} and the following identity
$[\D, D_{\b}{}^{I}] = \frac{2}{3} \pa_{\b \g} D^{\g I}$.
 
 A linearised gauge-invariant action for the $\cN=3$ superconformal gravitino multiplet 
 is fixed up to an overall constant. We propose the following action:
\bea
S [  H] = \hf
   \int \rd^{3|6}z \, H W (H) 
   \label{514}
\eea
to describe the dynamics of the superconformal gravitino multiplet. 


\section{${\cal N}=4$ supersymmetry}

In this section we introduce ${\cal N}=4$ superprojectors and construct the superconformal field strength $W_{\a(n)}(H)$  following  approaches similar to those developed in section \ref{section5}. An expression for the ${\cal N}=4$ super-Cotton tensor will also be presented. 


\subsection{Superprojectors}

The ${\cal N}=4$ transverse projector is given by 
\bea
\P_{\a}{}^{\b} &=& -\frac{\ri \Delta}{24 \Box^2} D^{\b I} D_{\a}^{I}\big( 4 \Delta^2 - \Box \big)~.
\label{5.1}
\eea
The properties \eqref{5.2} can be proved using \eqref{1.4} and \eqref{2.22}, 
in conjunction with the following identity:
\bea
\Delta^5 &=& \frac{1}{4} \Big\{ 5 \Box \Delta^3 - \Box^2 \Delta\ \Big\}~. \label{5.3d}
\eea
Unlike in the ${\cal N}=3$ case \eqref{4.4},  we are now required to use the full expression of the projector $\P_{\a}{}^{\b}$ in order to prove \eqref{5.2a}.


\subsection{Linearised rank-$n$ super-Cotton tensor}

In accordance with \eqref{1.9}, the dimensions of the ${\cal N}=4$ gauge prepotential $H_{\a(n)}$ and its corresponding field strength $W_{\a(n)}(H)$ are $(-2-\frac{n}{2})$ and $(1+\frac{n}{2})$, respectively. 
The dimensional analysis \eqref{1.9} and the conditions \eqref{1.7} and \eqref{1.8} imply that the field strength $W_{\a(n)}(H)$ is fixed, modulo an overall constant, in the form
\bea
W_{\a(n)} \big(H\big) =  \D^{n+3} \, \P^{\perp}_{[n]}H_{\a(n)}~,
\label{W4}
\eea
which is a special case of \eqref{2.16}.

Making use of \eqref{5.1} and the property \eqref{1.4}, the expression \eqref{W4} turns into
\bea
W_{\a(n)}(H) = \Big(-\frac{\ri}{24 \Box^2}\Big)^n \Delta^{2(n+1)} 
\Delta \big(4 \Delta^2 -\Box \big)^n 
D^{\b_1 I_1}D^{I_1}_{\a_1} \dots D^{\b_n I_n} D^{I_n}_{\a_n} H_{\b_1 \dots \b_n}~.
\label{5.6}
\eea
The expression \eqref{5.6} contains $\Box^{2n}$ in the denominator. However, it is possible to simplify it further using the following observations. First, equation \eqref{5.3d} leads to the following relation
\bea
\Delta (4 \Delta^2 - \Box)^n = (3\Box)^{n-1} \Delta (4 \Delta^2 - \Box)~, \qquad n=1,2,\dots~ 
\label{5.7}
\eea
Next, equation \eqref{1.3a} implies that for every transverse superfield $\J_{\a}$, we have
\bea
\Delta^2 \J_{\a} = \Box \J_{\a}~.
\label{5.8}
\eea
As a result, one may show that it is possible to cancel $\Box^{2n}$ in the denominator of \eqref{5.6} and thus arrive at
\bea
W_{\a(n)}(H) = \frac{1}{3}\Big(-\frac{\ri}{8} \Big)^n \D \big(4 \Delta^2 - \Box\big) 
D^{\b_1 I_1}D^{I_1}_{\a_1} \dots D^{\b_n I_n} D^{I_n}_{\a_n} H_{\b_1 \dots \b_n}~.
\label{5.9}
\eea
The superconformal field strength \eqref{W4}, or equivalently \eqref{5.9},
can be called  the rank-$n$ super-Cotton tensor. 

As a direct consequence of \eqref{5.3d} and \eqref{5.7}, we deduce that the following operator\bea
\cP = \frac{\D^2}{3\Box^2} (4\D^2 - \Box)
\eea
 is a projector, $\cP^2 = \cP$.


\subsection{Linearised $\cN=4$ conformal supergravity}

In the ${\cal N}=4$ case, the  linearised super-Cotton tensor proves to be a real scalar superfield, which obeys the following condition
\bea
(D^{\a I} D^{J}_{\a} - 2 \ri \d^{IJ} \Delta)W =0~.
\label{5.10}
\eea
In accordance with \eqref{1.9}, both $W$ and $H$ are primary superfields 
of dimension 1 and $-2$, respectively.
It is worth pointing out that if we act with $D^{\b J}$ on both sides of equation \eqref{5.10} and making use of \eqref{1.1} along with 
$[\D, D_{\b}{}^{I}] = \hf \pa_{\b \g} D^{\g I}$,
the resulting equation is 
\bea
D^{\a I} D^{\b J} D_{\a}^J \D \big( 4 \D^2 - \Box \big) =0~.
\eea
This is exactly the transversality condition of the projector $\P_{\a}{}^{\b}$ \eqref{5.2a}.
The super-Cotton tensor is given by 
\bea
W (H) = \frac{\D}{3}(4 \Delta^2 - \Box) H~,
\label{6.122}
\eea
which is the solution to \eqref{5.10}. 

We define the linearised action for $\cN=4$ conformal supergravity to be
\bea
S [  H] = \hf
   \int \rd^{3|8}z \, H W (H) ~.
\label{6.133}
\eea
It is invariant under the gauge transformations 
\bea
\d_\z H = \ri D^{\a I} D^{J}_{\a} \z^{IJ} ~, 
\qquad \z^{IJ} = \z^{JI}~, \quad \z^{II}=0~.
\eea


\section{${\cal N}=5$ supersymmetry}

The ${\cal N}=5$ transverse projection operator is given by 
\bea
\P_{\a}{}^{\b} &=& -\frac{\ri \Delta}{3840 \Box^3} D^{\b I} D_{\a}^{I}\big( 625 \Delta^4 - 250 \Box \D^2 +9 \Box^2 \big)~.
\label{6.1}
\eea
The projector properties \eqref{5.2} can be proved using eqs. \eqref{1.4} and \eqref{2.22}, 
in conjunction with the following identity:
\bea
\Delta^6 &=& \frac{1}{625} \Box \Big\{ 875 \Delta^4 -259 \Box \D^2 + 9 \Box^2 \Big\}~. \label{6.3d}
\eea
As in the ${\cal N}=3$ case, to prove the transversality condition \eqref{5.2a}, it suffices to show that the following relation holds
\bea
D^{\a J}\Big\{D^{\b I}D^{I}_{\a}\big( 625 \Delta^4 - 250 \Box \D^2 +9 \Box^2 \big)\Big\} =0~.
\eea

In accordance with \eqref{2.16}, the field strength $W_{\a(n)}(H)$ takes the following form
\bea
W_{\a(n)} \big(H\big) =  \D^{n+4} \, \P^{\perp}_{[n]}H_{\a(n)}~.
\label{W5}
\eea
Making use of \eqref{6.1} and the property \eqref{1.4}, the expression \eqref{W5} turns into
\bea
W_{\a(n)}(H) &=& \Big(-\frac{\ri}{3840\Box^3}\Big)^n \Delta^4 \Delta^{2n}\big(625 \Delta^4 -250 \Box \D^2 +9\Box^2 \big)^n 
\non\\
\qquad &&\times D^{\b_1 I}D^{I}_{\a_1} \dots D^{\b_n J} D^{J}_{\a_n} H_{\b_1 \dots \b_n}~.
\label{6.6}
\eea
It is possible to simplify \eqref{6.6} further. First, equation \eqref{6.3d} leads to the following relation 
\bea
\D^{2n} \big(625 \Delta^4 -250 \Box \D^2 +9\Box^2 \big)^n &=& 384^{n-1} \Box^{3(n-1)} \D^2 \big(625 \Delta^4 -250 \Box \D^2 +9\Box^2 \big)~~~
\label{6.7}
\eea
for $n \geq  1$.
As a result, using \eqref{6.7} and \eqref{5.8} one may show 
that it is possible to cancel $\Box^{3n}$ in the denominator of \eqref{6.6} 
to obtain
\bea
W_{\a(n)}(H) &=& \frac{1}{384} \Big( -\frac{\ri}{10} \Big)^{n} 
\big(625 \Delta^4 -250 \Box \D^2 +9\Box^2 \big) \non \\
&& \times 
D^{\b_1 I_1}D^{I_1}_{\a_1} \dots D^{\b_n I_n} D^{I_n}_{\a_n} H_{\b_1 \dots \b_n}~.
\label{7.77}
\eea

As a final observation, from equations \eqref{6.3d} and \eqref{6.7} we deduce that the following operator 
\bea
\cP = \frac{1}{384\Box^2} \big(625 \Delta^4 -250 \Box \D^2 +9\Box^2 \big)
\eea
is a projector, $\cP^2 =\cP$.


\section{${\cal N}=6$ supersymmetry}

The ${\cal N}=6$ transverse projection operator $\P_{\a}{}^{\b}$ is given by 
\bea
\P_{\a}{}^{\b} &=& -\frac{\ri \Delta}{480 \Box^3} D^{\b I} D_{\a}^{I}\big( 81 \Delta^4 - 45 \Box \D^2 +4 \Box^2 \big)~.
\label{7.1}
\eea
The superprojector properties \eqref{5.2} can be proved using \eqref{1.4} and \eqref{2.22}, 
in conjunction with the following identity:
\bea
\Delta^7 &=& \frac{1}{81} \Box \Big\{ 126 \Delta^5 -49 \Box \D^3 + 4 \Box^2 \D \Big\}~. \label{7.3d}
\eea

In accordance with \eqref{2.16}, the field strength $W_{\a(n)}(H)$ takes the following form
\bea
W_{\a(n)} \big(H\big) =  \D^{n+5} \, \P^{\perp}_{[n]}H_{\a(n)}~.
\label{W6}
\eea
Making use of \eqref{7.1} and the property \eqref{1.4}, the expression \eqref{W6} turns into
\bea
W_{\a(n)}(H) &=& \Big(-\frac{\ri}{480\Box^3}\Big)^n \Delta^{2n+5}\big(81 \Delta^4 -45 \Box \D^2 +4\Box^2 \big)^n 
\non\\
\qquad &&\times D^{\b_1 I_1}D^{I_1}_{\a_1} \dots D^{\b_n I_n} D^{I_n}_{\a_n} H_{\b_1 \dots \b_n}~.
\label{7.6}
\eea
It is possible to simplify \eqref{7.6} further. First, equation \eqref{7.3d} leads to the following relation
\bea
\Delta^2 \big(81 \Delta^4 -45 \Box \D^2 +4\Box^2 \big)^n = 40^{n-1} \Box^{2(n-1)} \D^2 \big(81 \Delta^4 -45 \Box \D^2 +4\Box^2 \big)~.
\label{7.7}
\eea
As a result, using \eqref{7.7} and \eqref{5.8} one may show that it is possible to cancel $\Box^{3n}$ in the denominator of \eqref{7.6} and thus arrive at
\bea
W_{\a(n)}(H) = \frac{1}{40} \Big(-\frac{\ri}{12} \Big)^n \D \big(81 \Delta^4 -45 \Box \D^2 +4\Box^2 \big) D^{\b_1 I_1}D^{I_1}_{\a_1} \dots D^{\b_n I_n} D^{I_n}_{\a_n} H_{\b_1 \dots \b_n}~.
~~~
\label{8.66}
\eea

Finally, let us point out that \eqref{7.7} allows us to show that the following operator
\bea
\cP = \frac{\D^2}{40 \Box^3}\big(81 \Delta^4 -45 \Box \D^2 +4\Box^2 \big)
\eea
is a projector, $\cP^2 = \cP$.


\section{Conclusion} 

In this paper we have presented a universal approach to construct linearised gauge-invariant  actions
for  higher-spin ${\cal N}$-extended superconformal gravity
in terms of the unconstrained prepotentials $H_{\alpha (n)}$,  $n>0$. 
These superconformal actions have the form \eqref{action}.
Our method was based on the use of the transverse superprojectors $ \P^{\perp}_{[n]}$, 
which exist for arbitrary $\cN$ and  have been explicitly
constructed   in this paper for $1\leq \cN\leq 6$.
We have  demonstrated that
the rank-$n$ super-Cotton tensor $W_{\alpha (n)}$ is given  in terms of the prepotential
$H_{\alpha (n)}$ by the universal expression \eqref{2.16}, 
and $W_{\a(n)}$ has been computed explicitly for $1\leq \cN\leq 6$.
In particular, a new expression \eqref{W2.2} 
for the rank-$n$ super-Cotton tensor in the case of $\cN=2$ supersymmetry has been obtained. 
This expression is much simpler than the one originally given in \cite{KO}. 

The rank-$n$ super-Cotton tensors $W_{\alpha (n)}$ for $3\leq \cN \leq 6$ 
were derived in this paper for the first time. 
The corresponding results 
are given by eqs. \eqref{4.10},   \eqref{5.9}, \eqref{7.77} and  \eqref{8.66}, respectively. 
The linearised super-Cotton tensor of ${\cal N}=4$ conformal supergravity 
requires special attention, since it is a scalar superfield $W$.
It has  also been computed in this paper for the first time and is given by eq. 
\eqref{6.122}. Making use of $W$ allowed us to construct the linearised action 
for $\cN=4$ conformal supergravity, which is given by eq. \eqref{6.133}. 
In the $\cN=3$ case, we also constructed the gauge-invariant action \eqref{514} 
which describes the dynamics of the superconformal gravitino multiplet.

In the case of conformal supergravity with  ${\cal N} \geq 5$, 
the super-Cotton tensor has a different tensorial form than that of $W_{\a(n)}$, see eq. \eqref{2.666}.
It cannot be directly obtained from our results and will be studied elsewhere.

A natural direction for future work is to extend the superconformal actions under study beyond the linearised level. To start with, one can consider the action \eqref{action}
for values of $n=3,2,1$ corresponding to  the linearised ${\cal N}$-extended conformal supergravity with $\cN=1,2,3$, respectively,  
and develop the Noether procedure to construct the higher-order terms in $H$. 
From the previous work \cite{BKNT-M2,KNT-M} we know that  
${\cal N}$-extended conformal supergravity does exist for $1 \leq \cN \leq 6$ 
as a nonlinear theory, however it is formulated  in terms of {\it constrained} supergeometry. 
This means that the Noether procedure should lead to the full nonlinear theory, now
formulated in terms of an {\it unconstrained} prepotential. A more ambitious generalisation is to attempt to apply the same technique to the higher-spin 
theories corresponding to greater values of $n$. Hence, our approach might be a useful laboratory to study vertices for higher spin superfields. 

Our analysis in $\cN$-extended Minkowski superspace can be generalised to arbitrary 
conformally flat backgrounds by applying the approach advocated in \cite{KP2}.

Making use of the superprojectors  $\P^{\perp}_{[n]}$ naturally leads to 
a supersymmetric extension of the Fierz-Pauli equations \cite{FP}.
More specifically, by applying   the projection operator $\P^{\perp}_{[n]}$ 
to a superfield $\J_{\a(n)}$ that obeys the Klein-Gordon equation, 
$(\Box -m^2 ) \J_{\a(n)} =0$, yields the transverse superfield $\J^{\perp}_{\a(n)}=\P^{\perp}_{[n]}\J_{\a(n)}$, which 
obeys the $\cN$-extended super Fierz-Pauli equations
\bea
(\Box -m^2) \J^{\perp}_{\a(n)}=0~, \qquad
D^{\b I} \J^{\perp}_{\b \a_1 \dots \a_{n-1}} &=& 0~.
\label{9.1}
\eea
For $n>1$ the latter implies the ordinary conservation condition 
\bea
\pa^{\b\g} \J^{\perp}_{\b\g\a_1\dots \a_{n-2}} =0 ~.
\eea

The general solution of \eqref{9.1} describes two superhelicity states, see Appendix 
\ref{appendixB} for the definition of the $\cN$-extended superhelicity operator.
If we are interested in an on-shell  massive supermultiplet of a definite superhelicity, 
we must deal with the following superhelicity projection operators 
\bea
{\mathbb P}^\pm_{[n]}:= \hf \Big(\mathbbm{1} \pm \frac{\D}{\sqrt{\Box}}\Big) \P^{\perp}_{[n]}
\eea
with the property 
\bea
\D {\mathbb P}^\pm_{[n]} = \pm \sqrt{\Box} {\mathbb P}^\pm_{[n]}~.
\eea
The operators ${\mathbb P}^+_{[n]}$ and ${\mathbb P}^-_{[n]}$ are orthogonal projectors:
\bea
{\mathbb P}^+_{[n]}{\mathbb P}^-_{[n]}=0~, \qquad 
{\mathbb P}^+_{[n]}{\mathbb P}^+_{[n]}={\mathbb P}^+_{[n]}~, \qquad 
{\mathbb P}^-_{[n]}{\mathbb P}^-_{[n]}={\mathbb P}^-_{[n]}~.
\eea
Applying the projector ${\mathbb P}^+_{[n]} $ or ${\mathbb P}^-_{[n]} $ to 
 a superfield $\J_{\a(n)}$ that obeys the Klein-Gordon equation, 
$(\Box -m^2 ) \J_{\a(n)} =0$, we will end up with an on-shell supermultiplet of
a definite superhelicity. We conclude this paper by giving a general definition
of such supermultiplets.

For $n>0$  a massive on-shell $\cN$-extended superfield is defined by 
the conditions
\begin{subequations} \label{9.4}
\bea
D^{\b I} T_{\b \a_1 \dots \a_{n-1}} &=& 0 
~ , 
\label{9.4a} \\
\D  T_{\a_1 \dots \a_n} &=& m \s T_{\a_1 \dots \a_n}~, 
\qquad \s =\pm 1~,
\label{9.4b}
\eea
\end{subequations}
of which the former implies 
$\pa^{\b\g} T_{\b\g\a_1\dots \a_{n-2}} =0$ for $n>1$.
This definition generalises those given earlier in the $\cN=1$ and $\cN=2$ 
cases \cite{KNT-M15,KT,KO}.

 In conclusion, we propose an off-shell gauge-invariant model 
in which the equations of motion are equivalent to \eqref{9.4}. It is a deformation 
of the superconformal action \eqref{action} given by 
\bea
{S}^{(n|\cN)}_{\rm massive} [  H_{\a(n)}] \propto \frac{ \ri^n}{2} 
   \int \rd^{3|2\cN}z \, H^{\a(n)} \Big( \D - m \s \Big)
{W}_{\a(n)}\big(H\big) ~, \qquad n>0~.
\label{action-mass}
\eea
Its invariance under the gauge transformation \eqref{1.6} follows from \eqref{1.3b}.
This model is a generalisation of the following massive gauge-invariant higher-spin theories:
(i) the non-supersymmetric models in Minkowski space \cite{BHT,BKRTY,BKLFP}
and anti-de Sitter space AdS${}_3$  \cite{KP}; 
and (ii) the supersymmetric models in AdS${}_3$
with (1,0) \cite{KP} and (2,0) \cite{HK18} anti-de Sitter supersymmetry.\footnote{There 
exist two alternative gauge-invariant formulations, off-shell and on-shell, for
massive higher-spin supermultiplets. The off-shell formulations have been developed for the cases $\cN=1$ and $\cN=2$ and are given in terms of 
the topologically massive higher-spin actions proposed in \cite{KO,KT,KP,HKO}.
The  on-shell formulations for
massive higher-spin $\cN=1$ supermultiplets in 
${\mathbb R}^{2,1}$ and AdS${}_3$ were  developed in \cite{BSZ3,BSZ4}
by combining the massive bosonic 
and fermionic higher-spin actions described in \cite{BSZ1,BSZ2}.
The formulations given in \cite{BSZ1,BSZ2,BSZ3,BSZ4}
are based on the gauge-invariant approaches 
to  the dynamics of massive higher-spin fields, which were advocated by Zinoviev \cite{Zinoviev} and Metsaev \cite{Metsaev}.}

In $\cN$-extended supersymmetry,  the conformal supercurrent has the same multiplet structure as the super-Cotton tensor \cite{KNT-M}. The Bianchi identities
\eqref{WW2.2}--\eqref{WW2.5} naturally lead to two types of 
supermultiplets of conserved currents
in ${\mathbb M}^{3|2\cN}$. One of them corresponds to the family of 
 real symmetric rank-$n$ spinor superfields $J_{\a(n) }$, $n=1,2,\dots $, 
subject to the conservation condition \cite{BKNT-M2,KNT-M,NSU}
\bea
D^{\b I} J_{\b \a_1 \dots \a_{n-1} } =0~, \qquad n>0~.
\eea
The second type is described by a real scalar superfield subject to the constraint  
\cite{BKNT-M2,KNT-M,BKS1}
\bea
\Big(D^{\a I} D^{J}_{\a} - \frac{1}{\cN}  \d^{IJ} D^{\a K} D^K_\a \Big)J =0~, \qquad \cN>1~.
\label{9.9}
\eea
The two types of conserved current supermultiplets are related to each other via 
the procedure of superspace $\cN \to (\cN-1)$ reduction described in \cite{BKS1}.
Let us  split the Grassmann coordinates $\q^\a_I $
of $\cN$-extended Minkowski superspace ${\mathbb M}^{3|2\cN}$ into two subsets:
(i) the coordinates $\q^\a_{\hat I}$, with $\hat I = \1, \dots, \underline{\cN-1}$,
corresponding to $(\cN-1)$-extended Minkowski superspace
${\mathbb M}^{3|2(\cN -1)}$; and (ii) two additional coordinates $\q^\a_{\underline \cN}$.
The corresponding splitting of the spinor derivatives $D_\a^I$  is
$D_\a^{\hat I} $ and $D_\a^{\underline \cN}$. Given a superfield $V$ on ${\mathbb M}^{3|2\cN}$,
its projection to ${\mathbb M}^{3|2(\cN-1)}$  is defined by
$V| := V|_{\q_\cN =0} $. The current supermultiplet $J_{\a(n)}$ produces two independent 
real superfields in ${\mathbb M}^{3|2(\cN -1)}$, which are defined by the rule 
\bea
{\mathfrak J}_{\a(n)} := J_{\a(n)}|~, \qquad 
{\mathfrak J}_{\a_1 \dots \a_{n+1} } := \ri^{n+1}
D_{\a_1}^{\underline \cN} J_{\a_2 \dots \a_{n+1}}|
={\mathfrak J}_{(\a_1 \dots \a_{n+1} )} 
\eea
and obey the conservation equations 
\bea
D^{\b \hat I} {\mathfrak J}_{\b \a_1 \dots \a_{n-1} } =0~, \qquad 
D^{\b \hat I} {\mathfrak J}_{\b \a_1 \dots \a_{n} } =0~.
\eea
The scalar current supermultiplet $J$ produces two independent 
real superfields in ${\mathbb M}^{3|2(\cN -1)}$, which are defined by the rule 
\bea
{\mathfrak J} := J|~, \qquad 
{\mathfrak J}_{\a} := \ri
D_{\a}^{\underline \cN} J |~,
\eea
which obey  the conservation equations 
\bea
D^{\b \hat I} {\mathfrak J}_{\b } =0~, \qquad 
\Big(D^{\a \hat I} D^{\hat J}_{\a} - \frac{1}{\cN-1}  \d^{\hat I \hat J} D^{\a \hat K} D^{\hat K}_\a \Big)
{\mathfrak J} =0~.
\eea
This consideration shows that a current supermultiplet $J_{\a(n)}$ 
in ${\mathbb M}^{3|2\cN }$ can be generated via Grassmann dimensional reduction 
from a scalar current supermultiplet in ${\mathbb M}^{3|2(\cN +n)}$.

The results of this paper provide general expressions for identically conserved (higher-spin) current supermultiplets. Identically conserved supercurrents $J_{\a(n)}$ are given by 
eq. \eqref{2.16} in which $W_{\a(n)}$ has to be identified with $J_{\a(n)}$ and $H_{\a(n)}$ 
should be viewed as a local operator constructed in terms of the dynamical superfields 
and their derivatives. Expressions for the identically conserved supercurrent 
$J$ follow from 
the relations \eqref{512} and \eqref{6.122} in the cases of $\cN=3$ and $\cN=4 $
supersymmetry, respectively. The   identically conserved $\cN=2$ scalar current
supermultiplet is given by 
\bea
J \propto \D H~, \qquad \cN=2~, 
\label{9.14}
\eea
which is a real linear superfield. It is natural to use the name ``$\cN$-extended linear multiplet''
for the real linear superfield $J$ constrained by eq. \eqref{9.9}.
Finally, in the cases $\cN=5$ and $\cN=6 $
scalar current supermultiplet has the form
\bea
J & \propto & \big(625 \Delta^4 -250 \Box \D^2 +9\Box^2 \big)H ~, \qquad \cN = 5, 
\label{9.15}
\\
J & \propto &\D  \big(81 \Delta^4 -45 \Box \D^2 +4\Box^2 \big) H~, \qquad \cN=6~.
\label{9.16}
\eea
It is an instructive exercise to check that these expressions satisfy the constraint 
\eqref{9.9}.
One may check that the currents \eqref{9.14},  \eqref{9.15} and  \eqref{9.16}
are invariant under  gauge transformations of the form
\bea
\d_\z H = \ri D^{\a I} D^{J}_{\a} \z^{IJ} ~, 
\qquad \z^{IJ} = \z^{JI}~, \quad \z^{II}=0~.
\eea

\noindent
{\bf Acknowledgements:}\\
We are grateful to Darren Grasso for comments on the manuscript.
The work of DH is supported by the Jean Rogerson Postgraduate Scholarship and an Australian Government Research Training Program Scholarship at The University of Western Australia.
The work of JH is supported by an Australian Government Research 
Training Program (RTP) Scholarship.
The work of SMK is supported in part by the Australian 
Research Council, project No. DP160103633.

\appendix

\section{Superconformal primary multiplets}\label{appendixA}

The $\cN$-extended superconformal symmetry 
in three dimensions was studied in detail by Park \cite{Park}.
In this appendix our presentation follows \cite{KPT-MvU,BKS1}.
In $\cN$-extended Minkowski superspace ${\mathbb M}^{3|2\cN}$, 
superconformal transformations, $z^A \to z^A +\d z^A = z^A +\x^A(z)$, 
are generated by superconformal Killing vectors. By definition, a superconformal Killing vector 
\bea
\x  = \x^{a} (z) \, \pa_{ a} 
+  \x^{\a}_I(z) \, D_{\a}^I 
\eea
is a real vector field obeying the condition
$[\x, D_\a^I ] \propto D_\b^J$. This condition implies
\bea
[\x, D_\a^I ] = -(D^I_\a \x^\b_J) D^J_\b =\hf  \o_\a{}^\b (z)D_\b^I + \L^{IJ}(z) D_\a^J -\hf \s (z) D^I_\a~,
\label{master1}
\eea
where we have defined 
\begin{subequations}
\bea
\o_{\a\b} &:=& -\frac{2}{\cN} D^J_{(\a} \x_{\b)}^J =-\frac{1}{2} \pa^\g{}_{(\a} \x_{\b )\g} ~,
\label{Lorentz}\\
\L^{IJ} &:=& -2 D_\a^{ [I} \x^{J]\a}~, \label{rotation}\\
\s&:=& \frac{1}{\cN} D^I_\a \x^{\a I} =\frac{1}{3} \pa_a \x^a~.
\label{scale}
\eea
\end{subequations}
Here the parameters $\o_{\a \b} =\o_{\b \a}$, $\L^{IJ}=- \L^{JI}$ and $\s$ correspond
to  $z$-dependent Lorentz, ${\rm SO} (\cN )$ and scale transformations.
These transformation parameters are related to each other as follows:
\begin{subequations}
\bea
D^I_\a \o_{\b \g} &=& 2 \ve_{ \a ( \b} D^I_{ \g)} \s~,  \label{Lorentz-scale}\\
D^I_\a \L^{JK} &=& -2 \d^{ I [ J } D_\a^{ K] } \s~. \label{rotation-scale}
\eea
\end{subequations}

Let  $\Phi_{\cal A}^{\cal I}(z)$ be a superfield that transforms
in a
representation $T$ of the Lorentz group with respect to its index $\cA$
and in a representation $D$ of the $R$-symmetry group ${\rm SO}(\cN)$
with respect to the index  $\cI$.
Such a superfield is called primary of dimension $d$ if its
superconformal transformation law is
\bea
\delta\Phi_{\cal A}^{\cal I} =
-\xi\Phi_{\cal A}^{\cal I}-d \,\sigma \Phi_{\cal A}^{\cal I}
+\hf \o^{\alpha\beta} (M_{\alpha\beta})_{\cal A}{}^{\cal B}
\Phi_{\cal B}^{\cal I}
+\hf \Lambda^{IJ}(R^{IJ})^{\cal I}{}_{\cal J}\Phi_{\cal A}^{\cal J}~.~~
\eea
The Lorentz generator $M_{\a\b}=M_{\b\a}$ is defined to act on a
 spinor $\F_{\g}$ by the rule
\bea
M_{\a\b} \F_{\g} = 
\ve_{\g (\a} \F_{\b) \g}~.
\eea
The ${\rm SO}(\cN)$  generator $R^{IJ}$ acts on an ${\rm SO}(\cN)$-vector $V^K$ as
\bea
R^{IJ} V^K = 2 \d^{K[I} V^{J]}~.
\eea
Making use of this transformation law allows one to determine the dimensions in
\eqref{1.9}.


\section{Superhelicity}\label{appendixB}

In this appendix we demonstrate how the operator $\D$ defined by \eqref{1.2}
emerges in the framework of superhelicity.

Let $P_{a}$,  $J_{ab}= -J_{ba}$, $Q_\a$ 
 be the generators of the $\cN$-extended super-Poincar\'e group
 in three dimensions. Important for our discussion are the following graded commutation relations:
\begin{subequations}
 \bea
 \big[ P_{\a\b} , P_{\g \d} \big] &=& 0~,\\
\big\{  Q^I_\a ,Q^J_\b \big\} &=&2 \d^{IJ} P_{\a\b}~, \\
\big[ J_{\a\b} , Q^K_\g \big] &=& \ri \ve_{\g(\a} Q^K_{\b)} ~, \\
\big[ J_{\a\b} , P_{\g \d} \big] &=& \ri \ve_{\g(\a} P_{\b)\d} + \ri \ve_{\d(\a} P_{\b)\g}~.
\eea
\end{subequations}
 The supersymmetric extension of the Pauli-Lubanski scalar 
 ${\mathbb W}=  \hf \ve^{abc}P_a J_{bc} = -\hf P^{\a\b} J_{\a\b}$
  is the  superhelicity operator  \cite{MT}
\bea
{\mathbb Z}= {\mathbb W}-\frac{\ri}{8} Q^{\a I} Q_\a^I
~,
\label{super-helicity}
\eea
which commutes with the supercharges,
\bea
\big[ {\mathbb Z},Q^I_\a \big] =0~.
\eea
It is worth pointing out that the structure of 
 \eqref{super-helicity} is analogous to the superhelicity operator 
 in four-dimensional $\cN=1$ supersymmetry \cite{BK}.
Given an irreducible unitary representation of the  super-Poincar\'e group, 
the quantum numbers of  mass $m$ and superhelicity $\k$ 
are defined by 
\bea
 P^a P_a = -m^2 {\mathbbm 1} ~, \qquad {\mathbb Z}=m  \k {\mathbbm 1}~.
 \eea

For superfield representations of  the $\cN$-extended super-Poincar\'e group, 
the  infinitesimal super-Poincar\'e transformation of a tensor superfield $T$
(with suppressed indices) is given by 
\bea
 \d T =\ri (-b^a P_a +\hf \o^{ab} J_{ab} + \ri \e^{\a I}Q^I_\a) T
 =\ri \Big(\hf b^{\a\b} P_{\a\b}  +\hf \o^{\a \b} J_{\a \b} + \ri \e^{\a I} Q^I_\a \Big)T~,
 \eea
 where the generators of spacetime translations ($P_{\a\b}$), 
 Lorentz transformations ($J_{\a\b}$) and supersymmetry transformations 
 ($Q^I_\a$) are 
 \begin{subequations}
 \bea
 P_{\a\b}&=& -\ri \pa_{\a\b} ~,\qquad \pa_{\a\b}= (\g^a)_{\a\b} \pa_a~, \\
 J_{\a\b} &=& -\ri x^\g{}_{(\a} \pa_{\b)\g } 
 + \ri \q^I_{(\a} \pa^I_{\b)} 
 -\ri M_{\a\b}~,\\
 Q^I_\a &=& \pa^I_\a -\ri \q^{\b I} \pa_{\a\b} ~,\qquad \pa^I_\a =\frac{\pa}{\pa \q^{\a I}}~.
 \eea
 \end{subequations}
These expressions allow us to represent 
 the superhelicity operator \eqref{super-helicity}  in a
 manifestly supersymmetric form 
 \bea
 {\mathbb Z} =\hf \pa^{\a \b} M_{\a\b} +\frac{\cN}{4} \D~,
 \qquad \big[ {\mathbb Z} , D^I_\a \big]=0~,
 \eea
with the operator $\D$ defined by \eqref{1.2}. For completeness, 
we recall  the explicit form
of the spinor covariant derivative 
\bea
 D^I_\a = \pa^I_\a +\ri \q^{\b I} \pa_{\a\b} ~.
 \eea

Consider a massive on-shell superfield of the type \eqref{9.4}.
Its superhelicity is equal to 
\bea
\k =\hf \Big( n + \frac 12 \cN \Big) \s~.
\eea 
 The independent component fields of $T_{\a(n)}$ 
may be chosen as 
\bea
\F^{I_1 \dots I_k}_{\a_1 \dots \a_{n+k} } (x) := 
\ri^{ nk + \hf k(k+1) }D^{[I_1}_{\a_1} \dots D^{I_k]}_{\a_k} 
T_{\a_{k+1} \dots \a_{n+k} } \Big|_{\q=0}~, \qquad 0 \leq k \leq \cN~.
\eea
Each of these fields is completely symmetric in its spinor indices,
$\F^{I_1 \dots I_k}_{\a_1 \dots \a_{n+k} } =\F^{I_1 \dots I_k}_{(\a_1 \dots \a_{n+k} )} $,
and proves to  be transverse, 
\bea
\pa^{\b\g} \F^{I_1 \dots I_k}_{\a_1 \dots \a_{n+k-2} \b\g } =0~, \qquad n+k>1~.
\eea
Its helicity is equal to $\hf \Big( n  +k \Big) \s$.

\begin{footnotesize}

\end{footnotesize}


\end{document}